\documentclass[apj,twocolumn]{emulateapj}
\usepackage{apjfonts}
\usepackage[pagebackref=false,colorlinks=true,citecolor=blue,linkcolor=blue,breaklinks=true,bookmarks=true]{hyperref}
\usepackage{amsmath}
\usepackage{lineno}
\usepackage{CJKutf8}

\newcommand{\reff}{$R_{\rm eff}$}
\newcommand{\ewha}{EW(H$\alpha$)}
\newcommand{\logm}{log($M/M_{\odot}$)}
\defcitealias{Fu:2018}{Paper I} 

\begin{document}
\begin{CJK*}{UTF8}{gbsn}
\title{SDSS-IV MaNGA: The Radial Profile of Enhanced Star Formation in Close Galaxy Pairs}

\author{
Joshua~L.~Steffen\altaffilmark{1}, Hai~Fu\altaffilmark{1,2}, J.~M.~Comerford\altaffilmark{3}, Y.~Sophia~Dai (戴昱)\altaffilmark{4}, Shuai~Feng\altaffilmark{5,6,7}, Arran~C.~Gross\altaffilmark{1}, and Rui~Xue\altaffilmark{1}
}

\altaffiltext{1}{Department of Physics \& Astronomy, University of Iowa, Iowa City, IA 52242}
\altaffiltext{2}{Insititute for Astronomy, University of Hawaii, Honolulu, HI 96822}
\altaffiltext{3}{Department of Astrophysical and Planetary Sciences, University of Colorado, Boulder, CO 80309}
\altaffiltext{4}{Chinese Academy of Sciences South America Center for Astronomy (CASSACA)/National Astronomical Observatories of China (NAOC), 20A Datun Road, Beijing 100012, China}
\altaffiltext{5}{Department of Physics, Hebei Normal University, 20 South Erhuan Road, Shijiazhuang, 050024, China}
\altaffiltext{6}{Key Laboratory for Research in Galaxies and Cosmology, Shanghai Astronomical Observatory, Chinese Academy of Sciences, 80 Nandan Road, Shanghai 200030, China}
\altaffiltext{7}{University of the Chinese Academy of Sciences, No.19A Yuquan Road, Beijing 100049, China}

\begin{abstract}
We compare the radial profiles of the specific star formation rate (sSFR) in a sample of 169 star-forming galaxies in close pairs with those of mass-matched control galaxies in the SDSS-IV MaNGA survey. We find that the sSFR is centrally enhanced (within one effective radius) in interacting galaxies by $\sim$0.3 dex and that there is a weak sSFR suppression in the outskirts of the galaxies of $\sim$0.1 dex. We stack the differences profiles for galaxies in five stellar mass bins between \logm\ $=$ 9.0$-$11.5 and find that the sSFR enhancement has no dependence on the stellar mass. The same result is obtained when the comparison galaxies are matched to each paired galaxy in both stellar mass and redshift. In addition, we find that that the sSFR enhancement is elevated in pairs with nearly equal masses and closer projected separations, in agreement with previous work based on single-fiber spectroscopy. We also find that the sSFR offsets in the outskirts of the paired galaxies are dependent on whether the galaxy is the more massive or less massive companion in the pair. The more massive companion experiences zero to a positive sSFR enhancement while the less massive companion experiences sSFR suppression in their outskirts. Our results illustrate the complex tidal effects on star formation in closely paired galaxies.
\end{abstract}


\section{Introduction}\label{sec:intro}

In the $\Lambda$CDM model, galaxy evolution is a hierarchical process. In this model, massive galaxies are the product of several past merger events of smaller galaxies. In fact, cosmological hydrodynamical simulations have shown that repeated merger events may be responsible for as much as $\sim$60\% of stellar mass in massive galaxies like M87 \citep[e.g.,][]{Rodriguez-Gomez:2016,Pillepich:2018}. As the galaxies undergo the merging process, the gas within the galaxies are subjected to gravitational torques which perturb the morphology of the galaxies.

The internal dynamics of these interacting galaxies were first modeled in the seminal work, \citet{Toomre:1972}. Since then, hydrodynamical simulations have expanded upon the N-body simulations of \citet{Toomre:1972} by modeling gas-dynamics within the galaxies. These simulations show the process by which barred structures develop within the disks of the interacting galaxies due to the tidal torques between them \citep{Barnes:1991}. As the bars form, the gases within the galaxy's disk lose angular momentum and get funneled into the centers of the galaxies. 

When the gas-inflows impact upon the gases in the nucleus of the galaxy, a burst of new star formation is triggered \citep{Barnes:1996, Mihos:1996}. These gas inflows will also bring metal-poor gases from the disk into the center of the galaxy which can dilute the central metallicity \citep{Rupke:2010, Perez:2011, Scudder:2012}. The gas-inflows may also be able to reach into the very center of the galaxy and trigger an episode of supermassive black hole (SMBH) accretion \citep{Capelo:2017}. 

Interaction induced star formation was first seen observationally in the bluer colors of peculiar galaxies in \citet{Larson:1978}. Similar observations have has also been shown in more recent works using the single-fiber spectroscopic survey, SDSS (Sloan Digital Sky Survey) \citep{Ellison:2008, Li:2008, Scudder:2012, Patton:2013, Bustamante:2020}. From these previous works it has been shown that the strength of the star formation enhancement in the centers of paired galaxies is dependent on the stellar mass of the pairs \citep{Li:2008}, the projected separation between the pairs \citep{Ellison:2008, Li:2008, Scudder:2012}, and the mass ratio between the pairs \citep{Ellison:2008}. 

The previously mentioned works using SDSS were restricted to studying the centers of the paired galaxies through 1-1.5\arcsec-radius optical fibers. With the recent large integral field spectroscopic (IFS) surveys, interacting galaxies can now be studied with unprecedented spatial detail. These surveys allow us to study the centers of merging galaxies more rigorously since apertures can be set to the physical scale of the galaxies instead of being bound by a fixed sky aperture. These IFS surveys will also allow us to see the extent of the centrally induced star formation and to see how the star formation in the disks of the galaxies are affected. 

Indeed, \citet{Barrera-Ballesteros:2015} used the CALIFA (Calar Alto Legacy Integral Field Area) survey to study a sample of 103 paired galaxies by varying the size of the aperture through which the \ewha\ is extracted. In this study, they found a moderate enhancement to the sSFR in the centers of paired galaxies and a moderate suppression to the sSFR in outskirts of the paired galaxies. 

\citet{Pan:2019} used the SDSS-IV MaNGA survey to study radial profiles of a sample of 205 paired galaxies. The enhancement to the sSFR was shown to be the strongest in the centers of the paired galaxies. This central enhancement linearly fell with increasing galactocentric radii; however, a moderate enhancement to the sSFR remains in the outskirts of the galaxies. \citet{Pan:2019} further studied the paired galaxies as a function of merger stage, from well separated pairs to post-merger galaxies. Across the different merger stages, the sSFR enhancement was greatest in close pairs with tidal features and in post-merger galaxies. This was in agreement with previous hydrodynamical simulations which showed that a burst of star formation is triggered after the first pericenter and as the two galaxies begin to coalesce \citep{Scudder:2012}. 

The radial profile of sSFR in post-merger galaxies has also been studied with the MaNGA survey by \citet{Thorp:2019}. The post-merger galaxies were shown to have a strong enhancement to the sSFR in their centers as well as a moderate enhancement in their outskirts. 

Where previous studies on the radial profile of the sSFR offsets in paired galaxies have focused on studying the profiles as a function of interaction stage, we will focus on the radial profile as a function of the stellar mass, projected separation, and mass ratio. As mentioned previously, these parameters have been covered by studies restricted to the nuclear region of the paired galaxies. With the MaNGA survey, we will be able to expand upon these previous studies in greater spatial detail. We will be able to analyze how these three parameters affect both the level of the sSFR offsets in the centers of the paired galaxies and the offsets in the outskirts of the galaxies. We will also be able to study whether any of the three parameters influence the gradient of the sSFR enhancement profiles or if the gradient is preserved between different configurations. 

In our previous work using the MaNGA data included in the 14th Public Data Release \citep[DR14;][]{Abolfathi:2018}, we built a sample of close galaxy pairs where both components of the pair were contained within the field of view of a single integral field unit \citep[][hereafter \citetalias{Fu:2018}]{Fu:2018}. We found that approximately 5.7\% of the MaNGA galaxies have a companion galaxy contained within the field-of-view of a single IFU. In this work, we update this sample and supplement it with a sample of companion galaxies identified outside the field-of-view of the MaNGA IFUs. 

This paper is organized as follows; in \S~\ref{sec:data} we will discuss the properties of the MaNGA survey along with the construction of our pair and control samples, in \S~\ref{sec:analysis} we will discuss how we measure star formation rates and how we build radial profiles of star formation, in \S~\ref{sec:results} we study the radial profiles as a function of stellar mass, projected separation, and the mass ratio, in \S~\ref{sec:disc} we compare our work against previous works, and in \S~\ref{sec:sum} we summarize the findings of the work. 
Throughout we adopt the $\Lambda$CDM cosmology with $\Omega_{\rm m}=0.3$, $\Omega_\Lambda=0.7$, and $h=0.7$. 

\section{Data and Samples}\label{sec:data}

\begin{figure*}
\centering
\includegraphics[width=\linewidth]{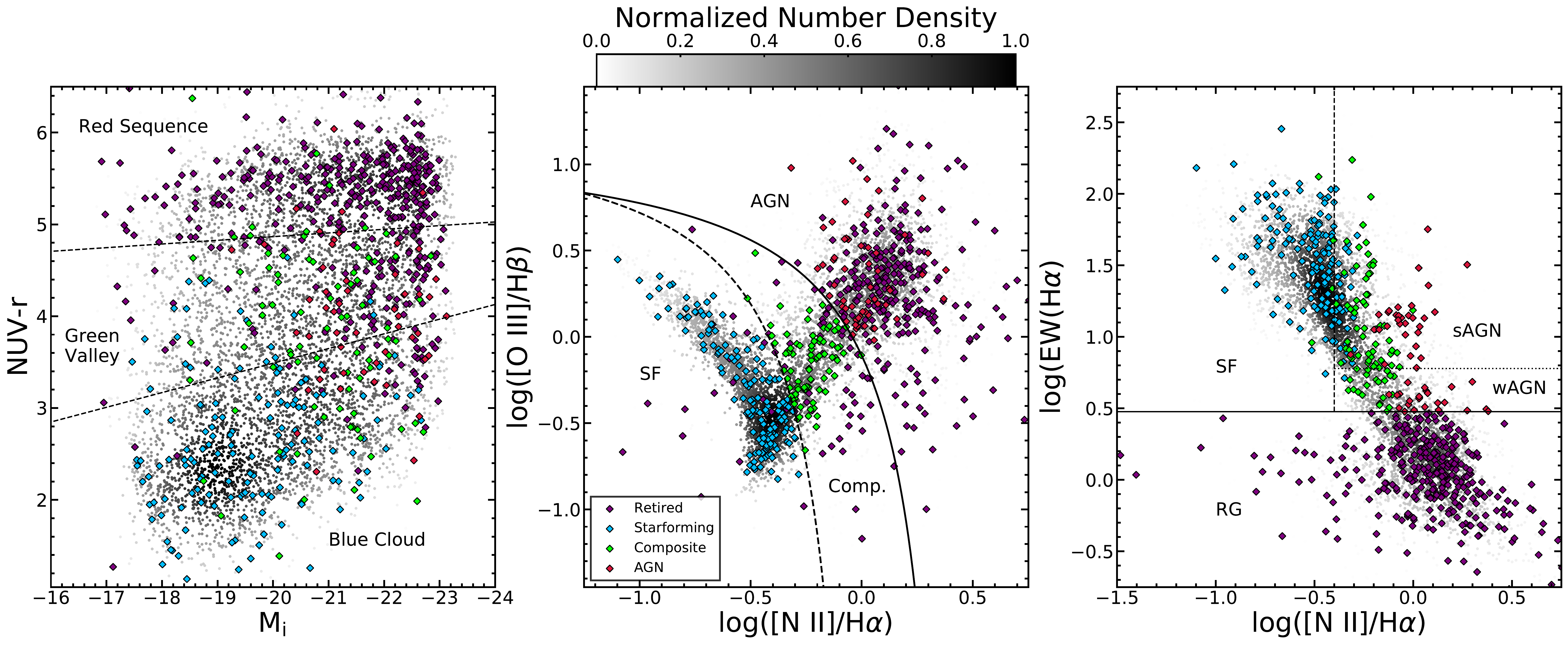}
\caption[]{(Left) Color-magnitude diagram for MaNGA galaxies. (Center) BPT diagram for the MaNGA galaxies. (Right) The WHAN diagram for the MaNGA galaxies. The grey circles represent the whole MaNGA sample and the color scale reflects the local density around each data point in this color-magnitude plane, as indicated by the color bar on the top. The {\it purple} diamonds represent the paired galaxies which are classified as retired (by the \ewha\ cut), the {\it blue} diamonds represent the star forming paired galaxies (by the BPT diagram), the {\it green} diamonds represent the composite star forming-AGN paired galaxies (by the BPT diagram), and the {\it Red} diamonds represent the paired galaxies containing an AGN (by the BPT diagram).}
\label{fig:cmd}
\end{figure*}

MaNGA is an IFS survey at the Apache Point Observatory (APO) which uses the SDSS (Sloan Digital Sky Survey) 2.5-meter telescope along with two dual-channel BOSS spectrographs \citep{Drory:2015}. MaNGA captures spectra through 17 integral field units (IFUs) with variable numbers of fibers: 19, 37, 61, 91, and 127 fibers covering 12.5\arcsec, 17.5\arcsec, 22.5\arcsec, 27.5\arcsec, and 32.5\arcsec on the sky respectively \citep{Law:2015}. MaNGA is an optical survey with a spectral coverage of 3600$-$10,300 \AA\ with a resolution of R $\sim$ 2000 and a PSF of 2.5\arcsec\ FWHM \citep{Bundy:2015}. 

The MaNGA survey targets galaxies from a subset of 41,154 galaxies in the NASA-Sloan Atlas (NSA v1\_0\_1; \url{http://www.nsatlas.org}) with a redshift range of 0.01 $< z <$ 0.15 and a luminosity range of -17.7 $<$ $\mathcal{M}_{\rm i}$ $<$ -24.0, where $\mathcal{M}_{\rm i}$ is the rest frame $i$-band magnitude within the survey's elliptical Petrosian apertures. MaNGA plans to cover 10,000 galaxies with a flat stellar mass distribution at two spatial coverages, 1.5 \reff\ and 2.5 \reff\ (where \reff\ is the radius which contains 50\% of the galaxy's total light). In this work we use the data from the 8th MaNGA Product Launch (MPL-8), which covers 6142 unique galaxies observed by July 3, 2018. 

\subsection{Spectral Fitting}\label{sec:spfit}

We use {\sc spfit} to model the MaNGA datacubes. The IDL package is publicly available\footnote{\url{https://github.com/fuhaiastro/spfit}} and was first used in our previous study of close galaxy pairs in MaNGA \citepalias{Fu:2018}. While data products from the MaNGA Data Analysis Pipeline \citep[DAP;][]{Belfiore:2019} and {\sc pipe3d} \citep{Sanchez:2016a, Sanchez:2016b} are available, {\sc spfit} allows us to combine spectra inside a given aperture before fitting the combined spectrum. The feature alone made it better suited for this project. Additional features of {\sc spfit} include: (1) simultaneously fitting emission lines and stellar continuum, (2) coadding spaxels with either a Voronoi tessellation or arbitrary polygons while accounting for covariances, (3) modeling asymmetric emission lines, broad AGN emission lines, AGN continuum, and various dust-extinction laws, and (4) utilizing multi-threading when processing a large number of datacubes.

Here we provide a brief description of the fitting procedure of {\sc spfit}. Except for broad-line AGN, {\sc spfit} models the observed spectrum as a superposition of emission lines and simple stellar populations (SSPs). The SSP library of {\sc miuscat} \citep{Vazdekis:2012} is matched to the MaNGA spectral resolution and is convolved with the line-of-sight velocity distribution (LOSVD). Both the LOSVD and the profile of the emission lines are parameterized as separate Gauss-Hermite series \citep{van-der-Marel:1993} to the fourth order.

For model optimization, we prefer a fast algorithm like the Levenberg-Marquardt nonlinear least-squares minimization algorithm implemented in {\sc mpfit} \citep{Markwardt:2009}. But for complex spectral models like ours, the success of the fitting routine relies on a good initial ``guess'' solution, which can be provided by the penalized pixel-fitting method \citep[pPXF;][]{Cappellari:2017}. The pPXF method is robust because it solves the weights of the templates with a linear algorithm \citep{Lawson:1974} {\it independently} from solving the Gauss-Hermite LOSVD with a nonlinear optimizer ({\sc mpfit}). The {\sc spfit} package implements a three-stage fitting procedure: First, it masks out spectral regions around emission lines from the input spectrum and use pPXF on the masked spectrum with SSP-only templates to obtain the initial ``best-fit'' parameters of the stellar continuum; Then, it subtracts the best-fit stellar continuum model from the spectrum and use pPXF to fit the residual emission-line-only spectrum with Gaussian emission-line templates to obtain the initial ``best-fit'' parameters of the emission lines; Finally, it uses {\sc mpfit} on the full input spectrum with the two sets of ``best-fit'' parameters from pPXF to simultaneously fit all of the parameters describing the emission lines and the stellar continuum. Saved in the final FITS file are the best-fit parameters and their uncertainties, including emission-line fluxes, equivalent widths, and kinematics and luminosity-weighted stellar masses, ages, metallicities, and kinematics. 

\subsection{Selecting Star Forming Galaxies}\label{sec:sf}

We classify galaxies in this survey as star forming galaxies by selecting galaxies in the blue cloud on the color-magnitude diagram (CMD). We show the CMD for the MaNGA survey and our pair sample in Figure \ref{fig:cmd} along with demarcation lines which separate the blue cloud, red sequence, and green valley. We established the demarcation lines by collapsing the CMD to a color histogram for each of the three regions. We then varied the slopes between the regions until we found the slopes which best fit the data. These demarcation lines are;

\begin{equation}\label{eq:blue}
NUV-r = 3.1682 - 0.16 (\mathcal{M}_{\rm i}+18)
\end{equation}
\begin{equation}\label{eq:red}
NUV-r = 4.7866 - 0.04 (\mathcal{M}_{\rm i}+18),
\end{equation}
Where $NUV-r$ is the color from SDSS's $k$-corrected absolute magnitude and $\mathcal{M}_{\rm i}$ is the $i$-band magnitude from the NSA catalog. 

We use the BPT diagnostic \citep{Baldwin:1981}, shown in the center of Figure \ref{fig:cmd}, to remove galaxies with possible AGN in their centers. Specifically, we use the emission line ratios, log([O~{\sc iii}]/H$\beta$) and log([N~{\sc ii}]/H$\alpha$), extracted from a 1.3~kpc aperture along with the maximum starburst line of \citet{Kewley:2001} as the demarcation between the star-forming branch and the AGN branch.

Based on the WHAN diagram, shown in the right panel of Figure \ref{fig:cmd}, we apply an \ewha\ $\ge$ 6\AA cut on galaxy spectra extracted from a 1.0~\reff\ aperture to ensure that all retired galaxies are removed from the sample \citep{Cid-Fernandes:2011}.

On top of the star formation cuts, we require that all galaxies are in either the Primary or Secondary MaNGA subsamples \citep{Wake:2017}, and that the stellar mass range of the galaxy sample is between \logm\ $=$ 9.0$-$11.5. We use masses calculated from the elliptical Petrosian apertures in the NSA catalog for our total stellar masses. 

We build two different pair samples in this work; the inside-IFU sample which contains paired galaxies where both galaxies are covered by a single MaNGA IFU and the outside-IFU sample where a MaNGA target galaxy is coupled with another galaxy found outside of its MaNGA IFU. 

\subsection{Inside-IFU Sample}\label{sec:inside}

\begin{table}
\begin{center}
\caption{Close Galaxy Pairs and Multiples in MaNGA IFUs}
\label{tab:sample}
\begin{tabular}{lc ccc}
\hline
\hline
Plate-IFU & Index & R.A. (J2000) & Decl. (J2000) & Redshift \\
          &           & (deg)        & (deg)         &     \\
\hline
 7443-12703&0&229.5255758&+42.7458538&0.04043\\
    \nodata&1&229.5265348&+42.7440666&0.04079\\
  7958-3701&0&257.5338400&+33.5989046&0.11015\\
    \nodata&1&257.5344200&+33.5984600&0.10920\\
    \nodata&2&257.5326566&+33.5979470&0.11031\\
  7960-3701&0&257.0857469&+31.7469109&0.10819\\
    \nodata&1&257.0850100&+31.7470300&0.10800\\
  7960-3702&0&258.2893316&+31.5820375&0.02961\\
    \nodata&1&258.2913540&+31.5812244&0.02943\\
  7962-3702&0&259.0637432&+28.0140065&0.10725\\
    \nodata&1&259.0639447&+28.0130439&0.10753\\
\hline
\end{tabular}
\end{center}
\tablecomments{
This table lists a total of 404 plate-IFUs that contain close galaxy pairs and multiples in the eighth MaNGA Product Launch (MPL-8). The second column lists the component index, where ``0'' indicates the primary target of the MaNGA observation. For each IFU, all components within $\pm$2000\,km~s$^{-1}$ of the primary are listed, sorted in ascending angular distance from the primary.\\
(This table is available in its entirety in a machine-readable form in the online journal. A portion is shown here for guidance regarding its form and content.)
}
\end{table}

To identify potential paired galaxies covered by individual IFUs, we start by overlaying SDSS photometric objects over each MaNGA fields of view. We manually inspect each MaNGA field, removing photometric objects which are over-deblended galaxy fragments and, very rarely, adding any objects missed in the SDSS photometric catalog. At this stage the object catalog includes foreground stars and foreground/background galaxies along with the potential paired galaxies. 

To select paired galaxies out of our object catalog, we inspect the spectra of each object. The spectra of the identified objects is extracted through a 1\arcsec\ circular aperture and fitted assuming the MaNGA target's redshift and then manually sorted into the following categories: ``good" galaxy spectra, broad-line AGN, foreground star, foreground/background galaxies, or poor S/N objects. The ``good" galaxy spectra are the objects whose spectra are well modeled by {\sc spfit} at the target galaxy's redshift, whether it is the target galaxy itself or a nearby companion galaxy. This means that the companion galaxy can be within approximately $\pm$2000 km s$^{-1}$ of the MaNGA target. We found 6573 ``good" objects, 57 broad-line AGN, 836 foreground stars, 319 foreground/background galaxies, and 1546 objects with poor S/N. 

Broad-line AGN comprise $\sim$0.9\% of MaNGA's galaxy sample where \citet{Lacerda:2020} found a fraction of $\sim$2.8\% from the CALIFA survey and \citet{Sanchez:2018} found a fraction of $\sim$1.33\% from the MaNGA's MPL-5 sample (which observed $\sim$2700 galaxies).  

Among the 6142 MaNGA IFUs considered here, 404 cover close galaxy pairs and multiples. The sample includes 327 pairs, 67 triplets, 7 quadruplets, and 1 quintuplet. Table\,\ref{tab:sample} lists the coordinates and redshifts of these galaxy components in pairs/multiples. When a MaNGA target has multiple companions galaxies, we chose the closest companion to define the pair's mass ratio and projected separation.

For this study, we further restrict the sample by setting a relative velocity cut of $\Delta v$ $<$ 500 km s$^{-1}$. Given the redshift range of the MaNGA sample and the size of MaNGA's IFUs, the maximum projected separation for a companion galaxy in the IFU is $\sim$40 kpc. Again, the galaxy sample is also restricted to galaxies with a stellar mass range of \logm\ $=$ 9.0$-$11.5 and galaxies in the Primary or Secondary MaNGA subsamples. We also require that the galaxies are classified as star forming as described in \S~\ref{sec:data}. These requirements reduce our sample down to 54 star forming MaNGA targets with inside-IFU companions. While we use the identified companions to select galaxy pairs, we only build radial profiles for the MaNGA target galaxies. We do this because the identified companions are not included in the NSA catalog which we use to build the radial profiles for the galaxies.

\subsection{Outside-IFU Sample}\label{sec:outside}

We supplement the inside-IFU sample with a set of pairs identified outside of the field of view of the MaNGA IFU. We select these outside-IFU pairs from the NSA catalog. We search for these external pairs by selecting objects with a projected separation from the MaNGA targets of $r_{\rm p}$ $<$ 50 kpc using the MaNGA target's redshift. We further use a relative velocity cut of $\Delta v$ $<$ 500 km s$^{-1}$ to remove projected galaxies from the selection. 

From the NSA catalog's 641,409 galaxies, we find 492 galaxies which are paired to MaNGA targets. After restricting the sample to SFGs, we have another 115 MaNGA targets with paired galaxies outside of the IFU. MaNGA targets which have both an inside-IFU and an outside-IFU pair are left to the inside-IFU sample.

\subsection{Control Sample}\label{sec:control}

To show how the population of galaxy pairs differs from isolated galaxies, we will compare our pair sample to a sample of control galaxies in the MaNGA survey. We build this control sample from the MaNGA target galaxies which have no spectroscopic companions within $r_{\rm p}$ $<$ 50 kpc and $\Delta v$ $<$ 500 km s$^{-1}$ either inside or outside of the IFU. This gives us a control sample of 1830 star forming control galaxies. 

It has been shown that the SFR in galaxies is dependent on both the stellar mass and the redshift of the galaxies \citep[e.g.,][]{Noeske:2007}. We will compare our interacting galaxies with control galaxies of a similar stellar mass, redshift, and radial size, to account for these other dependencies. To do this, we will use two different methods of pair - control comparison. 

In the first method, we simply control for the stellar mass between the pairs and controls. We split both the pair and control samples into five evenly spaced stellar mass bins over the range, \logm\ $=$ 9.0$-$11.5. The pairs are then compared to their respective controls within each stellar mass bins. While the redshift is not constrained in this method, we will show in \S~\ref{sec:mass-bin} that this method is sufficient to reveal the merger induced star formation. 

In the second method, we select a ``tailored'' control sample for each paired galaxy from the full sample of isolated galaxies. We select a subsample of 20 control galaxies for each paired galaxy. We first select all of the control galaxies which are within 0.1 dex in stellar mass, 0.025 in redshift, 20\% in effective radii, and within the same MaNGA subsample (e.g. Primary or Secondary) as the paired galaxy. With these requirements, most paired galaxies will find between 20$-$60 control galaxies. Since we want each paired galaxy to be treated in a similar manner, we randomly down-select the total number of acquired control galaxies to 20. A given control galaxy may be selected for multiple pairs by the pipeline; most of the control galaxies are used at least once and the average number of times a given control is reused is between 2$-$4 times. In the cases where a paired galaxy does not initially acquire 20 control galaxies we iteratively expand the stellar mass limit by 0.1 dex, the redshift by 0.025, and the \reff\ by 5\% until at least 20 control galaxies are found. 39 pairs required 1 extra iteration, 15 pairs required 2 extra iterations, 5 pairs required 3 extra iterations, and 2 pairs required 4 extra iterations. 

\section{Analysis Methods}\label{sec:analysis}

\subsection{Specific Star Formation Rate}
We calculate the star formation rates in the pair and control galaxies with the emission lines extracted with our spectra fitting code, {\sc spfit} (described in Section \ref{sec:spfit}). We correct the emission lines for reddening using the extinction curve from \citet{Cardelli:1989} with updated coefficients from \citet{ODonnell:1994}. The extinction is parameterized as $R_V$ $\equiv$ $A_V/E(B-V)$ $=$ 3.1, where we estimate the value of the $V$-band extinction, $A_V$, by comparing the H$\alpha$/H$\beta$ ratio to the expected value of 2.85 for Case-B recombination. 

We measure the SFR from the extinction corrected H$\alpha$ luminosity, $L_{{\rm H}\alpha}$. We use the SFR formula, Equation \ref{eq:sfr}, from \citet{Murphy:2011} which uses a Kroupa IMF, Solar metallicity, a constant SFR at an age of 100 Myr, and Case-B recombination: 
\begin{equation}\label{eq:sfr}
\frac{\rm{SFR}}{M_{\odot} \, \rm{yr^{-1}}} = \frac{L_{\rm H\alpha}}{1.86 \times 10^{41}\, \rm{erg\,s}^{-1}}.
\end{equation}

Since the stellar mass of a galaxy is not uniformly distributed within the galaxy, we normalize the SFR by the stellar mass in the same spaxel, $M$, giving us the specific star formation rate (sSFR):
\begin{equation}
{\rm sSFR} = \frac{\rm SFR}{M}.
\end{equation}
The local stellar masses used here is derived from {\sc spfit}'s best-fit stellar continuum. Utilizing sSFR, instead of SFR, allows us to compare regions of high mass surface density to regions of low mass surface density. 

We check our measurement of the specific star formation rate with the equivalent width of the H$\alpha$ line, \ewha, since it is a known proxy for the sSFR. This is useful as the sSFR is dependent on {\sc spfit}'s measurement of the stellar mass while \ewha\ is an observable. As we will show in Section \ref{sec:results}, the radial profiles of \ewha\ is consistent with radial profiles of the sSFR.

The signal to noise ratio of the data will decrease with wide galactocentric radii. MaNGA is designed to cover 1.5 \reff\ for galaxies in the Primary sample and 2.5 \reff\ in the secondary sample. These are not hard limits, data will exist beyond these radii, especially along the semi-minor axis of galaxy; however, the data beyond these limits may be unreliable due to low signal to noise. To control the quality of the used data, we only use spaxels with S/N $\ge$ 3 for the H$\alpha$ line. We also restrict the used spaxels to those with \ewha\ $\ge$ 6 \AA\ to remove retired regions from the survey \citep{Cid-Fernandes:2011}, which effectively sets -10.8 yr$^{-1}$ as the lower limit to the sSFR in our galaxies.

\subsection{Radial Profiles}\label{sec:radial}

In order to spatially characterize the star formation in the paired galaxies we build radial profiles of sSFR. First, the geometry of the galaxies needs to be defined. Specifically we will need the position angle, the inclination angle, and the effective radius of each of the MaNGA targets. We use the $r$-band elliptical Petrosian apertures from the NSA catalog and the $r$-band S\'ersic apertures from \citet{Simard:2011} to define the geometries of the galaxies. 

The NSA catalog has complete coverage over the MaNGA sample (since MaNGA selects its targets from this catalog); however, it tends to fail to properly fit paired galaxies with close on-sky separations. We found that the apertures from \citet{Simard:2011} work better for these close paired galaxies; however, the catalog does not completely cover the MaNGA sample. We use the NSA catalog for the outside-IFU pair sample because they are well separated on the sky. The \citet{Simard:2011} catalog is used for the inside-IFU sample. If the paired galaxy is not covered by \citet{Simard:2011}, we use the ellipse from the NSA catalog. 

We fit the geometry with apertures from both catalogs when available. When using the first method of pair-control comparison, where pairs and controls are grouped into stellar mass bins, we fit the control galaxies with the NSA apertures. When making the comparison with the second method, paired galaxies fitted with the NSA apertures are compared against controls fitted with the NSA apertures and paired galaxies fitted with the \citet{Simard:2011} apertures are compared against controls fitted with the \citet{Simard:2011} apertures. Further, when defining a galaxy's geometry with the \citet{Simard:2011} apertures we also use the masses given in the catalog for the total stellar mass of the galaxy. Note that although we carefully treat the two catalogs separately whenever possible, we find excellent agreement in the geometry parameters of control galaxies in both catalogs.

We calculate the inclination angle, {\it i}, of the galaxies using the major-to-minor axis ratios from the elliptical apertures;

\begin{equation}
{\rm cos^2}(i) = \frac{(b/a)^2 - q^2}{1 - q^2},
\end{equation}
Where $b/a$ is the major-to-minor axis ratio and $q$ is the intrinsic oblateness. We use the empirically determined oblateness of $q = 0.13$, from \citet{Giovanelli:1994}.

The inclination angle, along with the galaxy's position angle, is used to deproject the geometries of the galaxies. We use the 50\% half light radius ({\it i.e}, the effective radius, \reff) to scale the sizes of the galaxies. Doing this will allow us to compare galaxies of different sizes against each other.

Once the geometries of the galaxies are set, we can build azimuthally averaged radial profiles. The spaxels are binned into radius increments of 0.2 \reff\ from 0.0$-$2.6 \reff. Within each radius bin we take the median of the specific star formation rate. We build an azimuthally averaged radial profile for each MaNGA star forming galaxy and later stack and differentiate these profiles in the subsequent analysis. 

The MaNGA sample does not have a uniform spatial coverage, 63\% of the sample is designed to cover 1.5 \reff\ (the Primary+ subsample) and 37\% of the sample is designed to cover 2.5 \reff\ (the Secondary subsample). This means that there will be fewer selected spaxels beyond 1.5 \reff\ and the S/N of the selected spaxels will be lower than those within 1.5 \reff. We decide to still extend our radial profiles out to 2.5 \reff\ to make full use of the MaNGA data; however, we emphasize that the difference in the sampling of the data within 1.5 \reff\ and the data beyond 1.5 \reff\ may create artificial slopes in the data.

\section{Results}\label{sec:results}

\subsection{Star Formation Enhancement}

\begin{figure*}
\centering
\includegraphics[width=\linewidth]{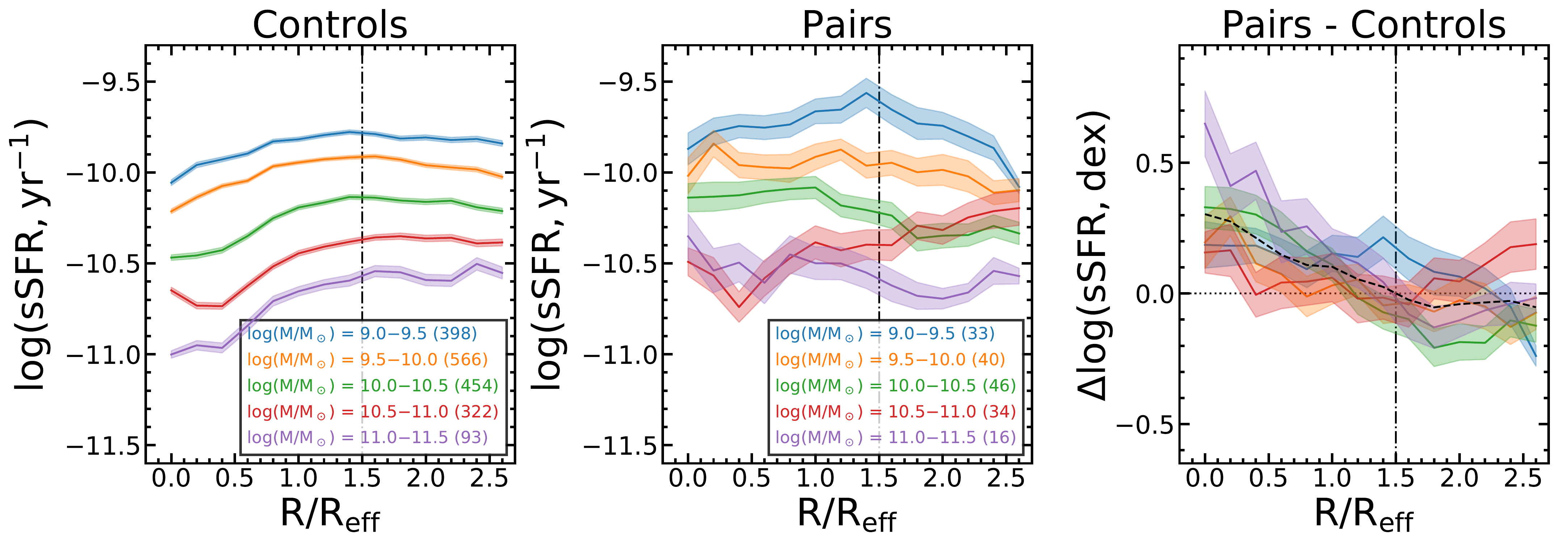}
\caption[]{The log(sSFR) as a function of galactocentric radius for control galaxies (Left) and galaxy pairs (Middle). The difference between the profiles of the paired galaxies and the control galaxies are shown in the Right panel. The dashed black profile represents the mean of the difference profiles. The colors of the profiles represent the mass range of the selected galaxies which is given in the legend along with the number of galaxies in that mass bin in parantheses. The highlighted region around the profiles represent the standard error of the mean of the data at the given radius interval. The vertical dash-dot line marks 1.5 \reff, beyond which the radial sampling is expected to fall.}
\label{fig:ssfr_prof}
\end{figure*}

\begin{figure*}
\centering
\includegraphics[width=\linewidth]{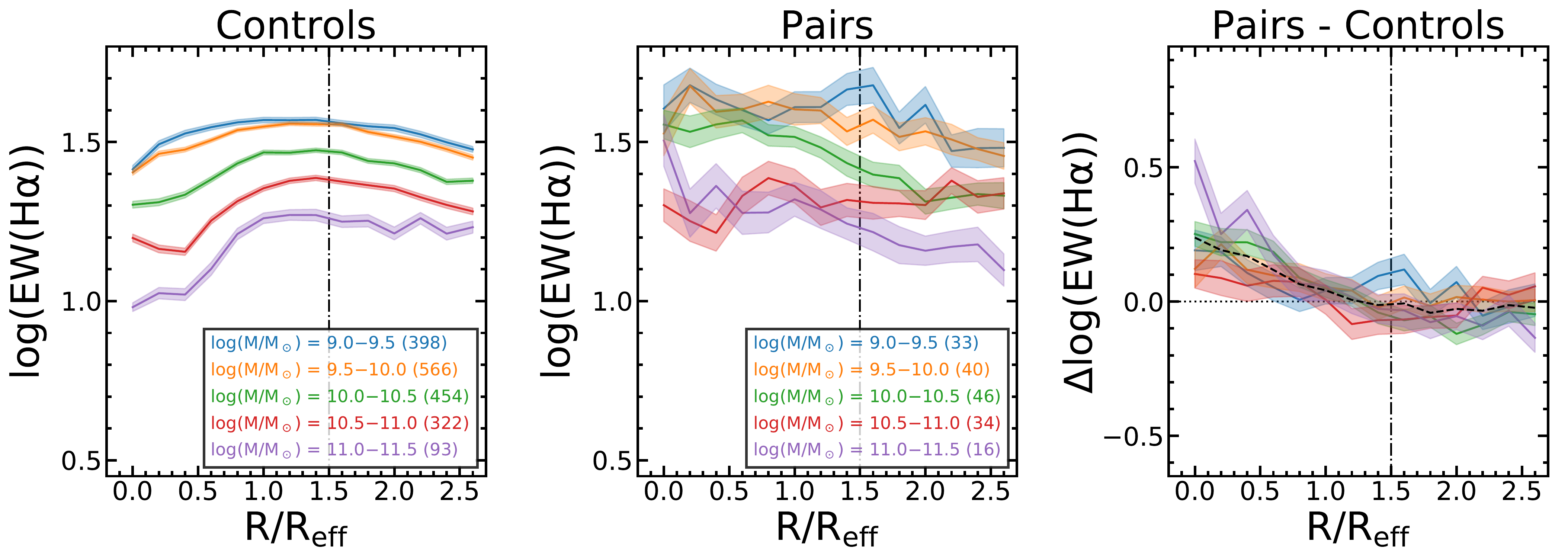}
\caption[]{Same as Figure \ref{fig:ssfr_prof} but for EW(H$\alpha$). }
\label{fig:ewha_prof}
\end{figure*}

With the individual log(sSFR) profiles built for each MaNGA galaxy, we now use two different methods to compare the profiles of paired galaxies against the profiles of control galaxies. In the first method, in \S~\ref{sec:mass-bin}, paired galaxies are compared to control galaxies within evenly spaced stellar mass-bins. In the second method, in \S~\ref{sec:tailored}, paired galaxies are compared to a subset of 20 control galaxies which are matched in stellar mass and redshift. 

\subsubsection{Mass-Binned Difference Profiles}\label{sec:mass-bin}

In the first method, paired and control galaxies are grouped into five evenly spaced stellar mass bins between \logm\ $=$ 9.0$-$11.5. Within each stellar mass bin, we create a median profile of log(sSFR) as a function of \reff-scaled galactocentric distance for both paired and control galaxies. The control profiles are constructed using all available control galaxies within each mass bin. The error associated with the ``stacked'' profile is the standard error of mean of the data at each radius bin. Here we define the standard error of the mean, $\sigma_{\overline{x}}$, as,
\begin{equation}
\sigma_{\overline{x}} = \frac{\sigma}{\sqrt{n}},
\end{equation}
where $\sigma$ is the standard deviation and $n$ is the sample size.

We show these stacked profiles for control galaxies in the left hand panel of Figure \ref{fig:ssfr_prof} and the stacked profiles for paired galaxies in the middle panel of Figure \ref{fig:ssfr_prof}. Star formation in the control galaxies are quenched in their centers with respect to their disks and the difference between the sSFR in their centers and their disks increases with stellar mass, consistent with previously published results using the MaNGA survey \citep{Belfiore:2018}. In contrast, the paired galaxies show flatter sSFR profiles where the level of the sSFR remains roughly consistent across their disks except for paired galaxies with stellar masses above \logm\ $=$ 10.5, which still show some central quenching. 

We then take the difference between the stacked profiles of the paired galaxies and the control galaxies, pair - control, in log space (this means that the difference profiles really represent a ratio between the pairs and controls in linear space). This gives us the difference profile, $\Delta$log(sSFR), which shows us where the sSFR is enhanced or suppressed (shown in the right hand panel of Figure \ref{fig:ssfr_prof}). Across all stellar mass bins, the sSFR of paired galaxies are centrally enhanced by $\sim$0.3$\pm$0.1~dex, which gradually falls to zero around $\sim$1.5 \reff. In the outskirts of the pairs beyond 1.5 \reff, the pairs feature lightly suppressed sSFR of 0.0$-$0.1 dex. All of the mass ranges except for the highest mass range, \logm\ $=$ 11.0$-$11.5, trend closely to the median profile. For the most massive galaxies, \logm\ $=$ 11.0$-$11.5, the enhancement to the sSFR within 0.5 \reff\ is significantly higher than the median profile (reaching 0.5$-$0.6 dex). As will be discussed in \S~\ref{sec:tailored}, we believe this elevated enhancement in the highest mass bin is driven by more compact mergers instead of stellar mass. 

In summary, Figure~\ref{fig:ssfr_prof} shows that the merger process is capable of rejuvenating star formation in the centers of these galaxies, resulting in almost flat sSFR radial profiles. In addition, the sSFR enhancement is mass independent -- star formation proceeds at $\sim$2$\times$ (0.3~dex) greater rates in galaxies with close companions than in more isolated galaxies, across a wide mass range between 9.0 $\leq$ \logm\ $\leq$ 11.5.

We repeat this analysis for the \ewha\ in Figure \ref{fig:ewha_prof}. We find that the \ewha\ profiles are in close agreement with the sSFR profiles, which shows that we can be confident of {\sc spfit}'s measurement of the sSFR. We do see that the pair-control offsets are lightly suppressed using the \ewha\ as the $\Delta$log(\ewha) is only enhanced by 0.25 dex where the $\Delta$log(sSFR) was enhanced by 0.30 dex. 

This method of stacking sSFR profiles by a single parameter (stellar mass) has the advantage of simplicity and large statistical samples. It has revealed the first order result of mass-independent, centrally enhanced star formation in close galaxy pairs. To proceed with exploring the dependency of $\Delta$sSFR on merger parameters such as separation and mass ratio, we will utilize a more sophisticated method of selecting control samples for individual paired galaxies based on stellar mass and redshift in the following subsections.  

\begin{figure*}
\centering
\includegraphics[width=0.8\linewidth]{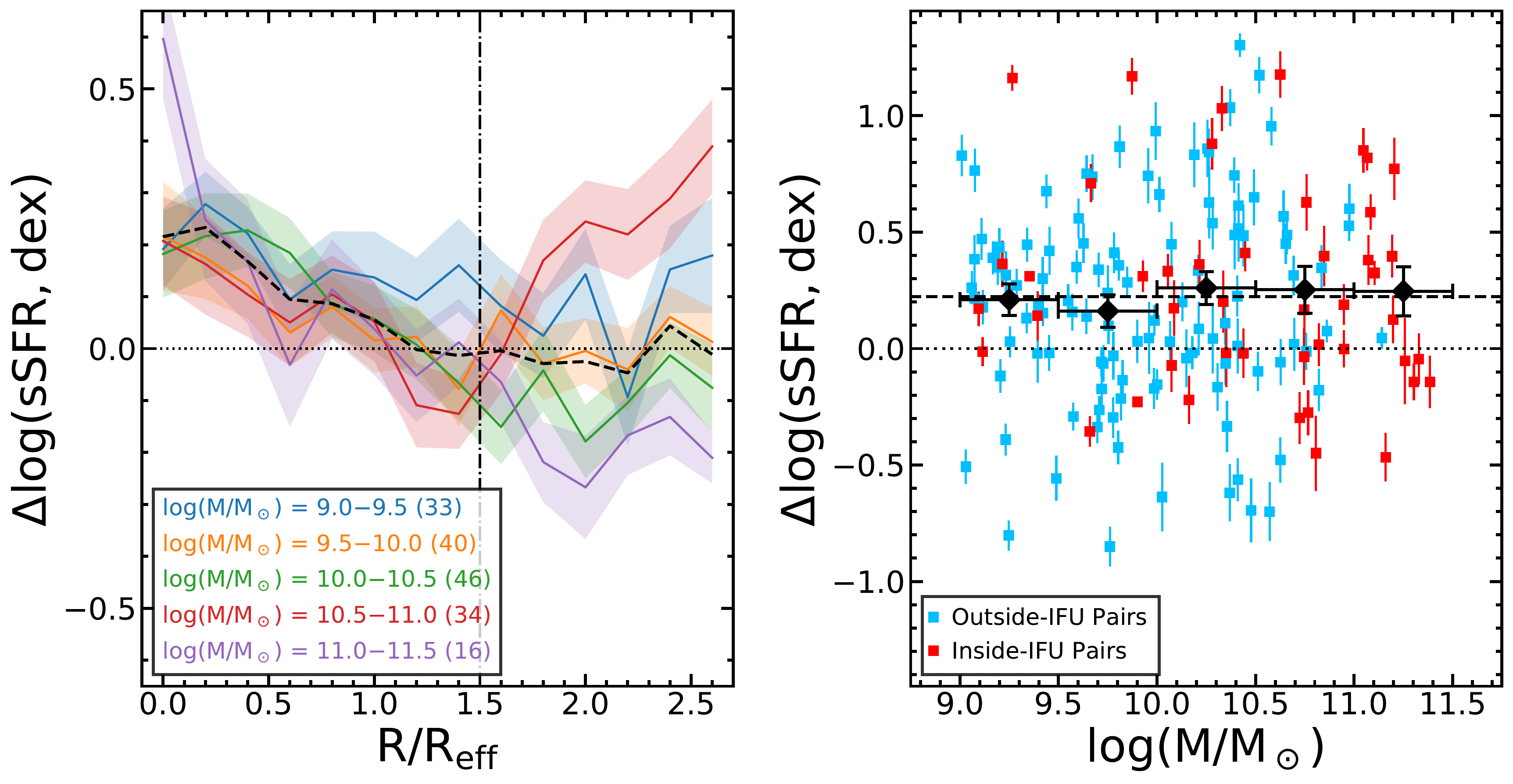}
\caption[]{The Left panel shows the $\Delta$log(sSFR) profiles where the difference profiles are constructed from the difference between the paired galaxy profiles and a set of 20 control galaxies. The profiles are split into five different stellar mass bins and the highlighted region about the profiles represent the standard error of mean of the profile. The black dashed line represents the mean profile between the four difference mass ranges. The number of paired galaxies in each mass range is given in the legend in parentheses. The Right panel shows the nuclear $\Delta$log(sSFR) extracted from a 0.5\reff\ aperture. The black squares are the mean values within a stellar mass bin (where the size of the bins are shown the the horizontal error bars). The vertical error bars on the black squares represent the standard deviation within the bin. The horizontal, dashed black line represents the median central enhancement of the pair sample. Galaxies in the outside-IFU (Blue) and inside-IFU (Red) samples are separately depicted. The vertical dash-dot line marks 1.5 \reff, beyond which the radial sampling is expected to fall. 
}
\label{fig:ssfr_mass}
\end{figure*}

\begin{figure}
\centering
\includegraphics[width=0.8\linewidth]{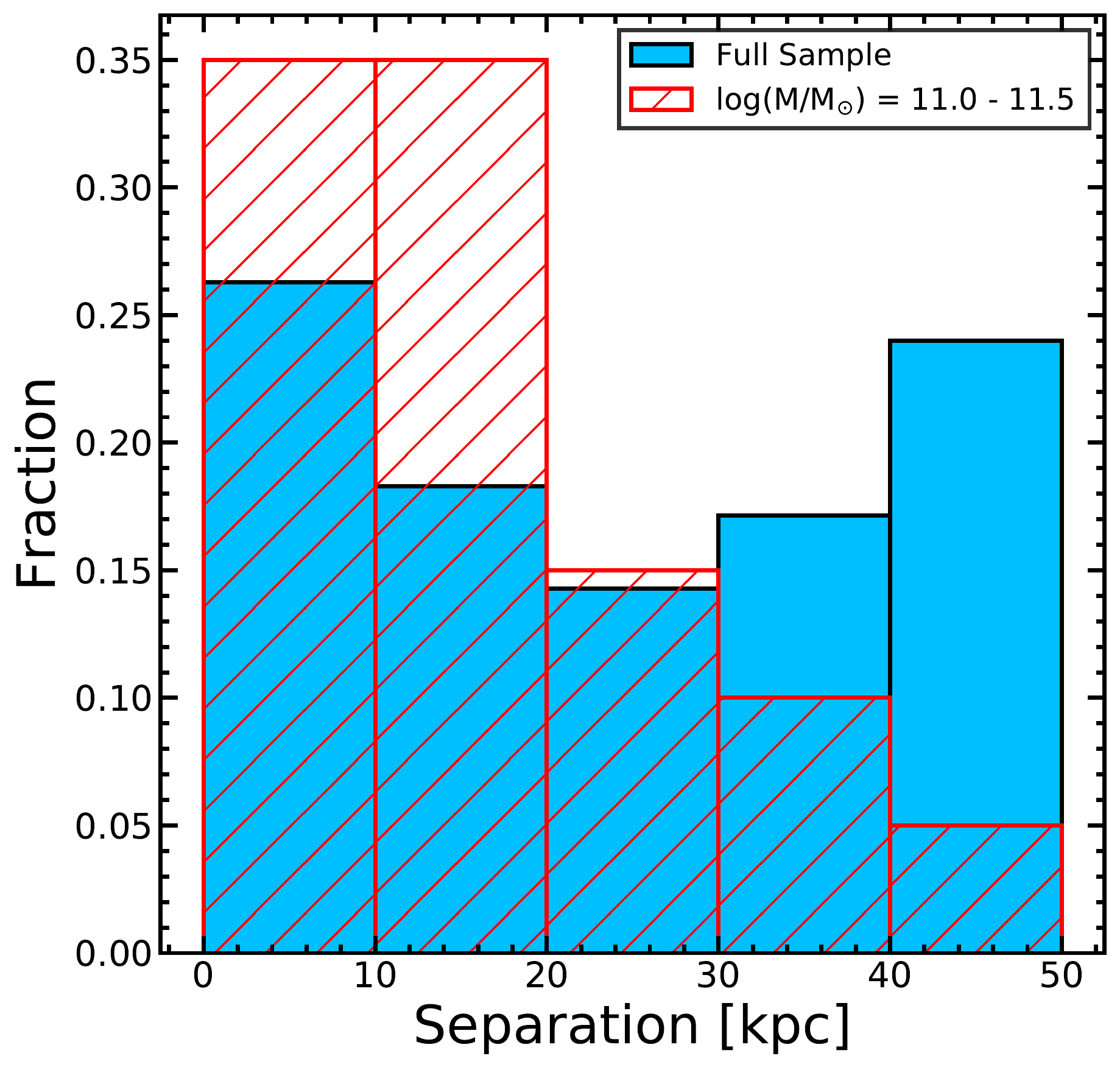}
\caption[]{The projected separation distribution of the whole sample (blue) against the projected separation distribution for the highest mass galaxies, \logm\ $=$ 11.0$-$11.5 (Red). }
\label{fig:sep_hist}
\end{figure}

\subsubsection{Tailored-Controls Difference Profiles}\label{sec:tailored}
\begin{figure*}
\centering
\includegraphics[width=0.8\linewidth]{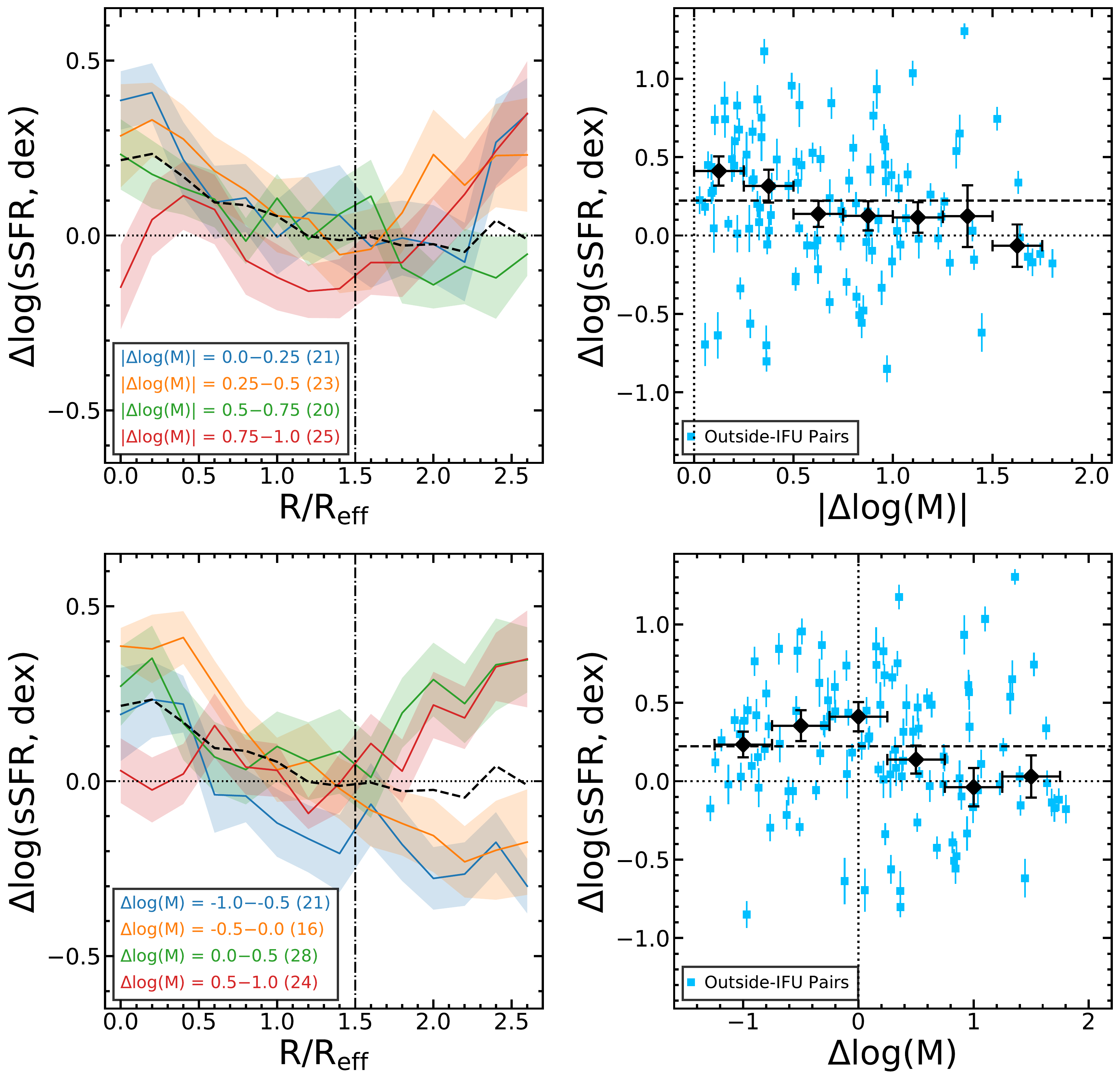}
\caption[]{Same as Figure \ref{fig:ssfr_mass} except the difference profiles are split by the mass ratio of the pair. The top row is the absolute value of the mass ratio and the bottom row is the mass ratio without taking the absolute value. Taking the mass ratio without the absolute value allows us to separately study the more massive companions of pairs from the less massive companions from pairs. The inside-IFU sample is left out of this analysis because we do not yet have reliable measurements of their total stellar mass ratios.}
\label{fig:ssfr_dm}
\end{figure*}
In the second method, we match each paired galaxy to a set of 20 control galaxies of similar stellar masses and redshifts, as described in \S~\ref{sec:control}. We then take the median of the azimuthally averaged profiles of the tailored control sample. Finally, we obtain the $\Delta\log$sSFR profile for each of the 169 paired galaxies by calculating the difference between its sSFR profile and the median profile of its control sample.

Before delving into merger parameters, we decided to stack the profiles by stellar mass to see how this method compares to the previous mass-binning method. We split the individual difference profiles into five evenly spaced stellar mass bins between \logm\ $=$ 9.0$-$11.5 and stack the difference profiles within each mass bin. This gives us five difference profiles covering five different stellar mass ranges shown in the left hand panel of Figure \ref{fig:ssfr_mass}. The errors of the stacked profiles are the standard error of the mean of the difference profiles within each bin.

To focus on the sSFR offsets in the centers of the galaxies, we also calculate the sSFR difference between each paired galaxy and its control sample within the central 0.5~\reff. The central sSFR is calculated for each galaxy by taking spaxels within 0.5 \reff\ and whose H$\alpha$ flux has S/N $\ge$ 3 and then taking the median value of sSFR of the selected spaxels. Once this is done we calculate the $\Delta$log(sSFR) using the same method as the $\Delta$log(sSFR) radial profiles are made. The central sSFR of a paired galaxy is compared against the sSFR of a set of 20 similar control galaxies, the same set of controls which were used for the $\Delta$log(sSFR) profiles, by taking the difference between the central sSFR of the paired galaxy and the median sSFR of the 20 selected control galaxies. We depict the central $\Delta$log(sSFR) as a function of stellar mass in the right hand panel of Figure \ref{fig:ssfr_mass}.

The difference profiles are shown to be centrally enhanced by $\sim$0.20$-$0.25 $\pm$ 0.1 dex. Similar to what was seen in \S~\ref{sec:mass-bin}, the level of the enhancement is independent of galaxy mass with the exception of the highest mass bin, \logm\ $=$ 11.0$-$11.5, which is 0.4 dex higher than the sample average. 

The central enhancement to the sSFR falls to zero around 1.2 \reff\ and the outskirts of the galaxies are lightly suppressed by 0.0$-$0.1 dex. Within individual mass ranges, there are some mass ranges which feature large enhancements to the sSFR at large radii while other mass ranges show a strong suppression to the sSFR. There appears to be no dependence on the stellar mass of the pair but, as we will show in the next section, \S~\ref{sec:dm}, these offsets at large radii may be dependent on the mass ratio between the galaxy pairs.

To understand why the profiles of high mass galaxies behave differently from the rest of the sample in Figures \ref{fig:ssfr_prof}, \ref{fig:ewha_prof}, and \ref{fig:ssfr_mass}, we compare the distribution of the projected separation for the whole sample and for galaxies in the highest mass bin in Figure \ref{fig:sep_hist}. The whole sample has a v-shaped separation distribution. This is the result of the combination of the two different samples, the inside-IFU and outside-IFU samples. The outside-IFU sample should find more pairs at wide projected separations due to the larger volume to find companions in. The separation distribution of the inside-IFU sample is limited by MaNGA's IFU sizes and the sample's redshift distribution; the maximum possible projected separation for a MaNGA target with $z$ $=$ 0.15 and a 127 spaxel IFU is about 40 kpc. 

The highest mass galaxies in the sample are at closer projected separations in comparison to the rest of the sample with 70\% of the highest mass galaxies being within 20 kpc.  The cause of this may be due to clustering effects where high mass galaxies tend to be at the centers of galaxy clusters \citep{Cooray:2002, Zehavi:2002}. As we will show in \S~\ref{sec:sep}, the central sSFR enhancement is strongest at small projected separation which means that higher sSFR in the centers of the massive galaxies may be driven by their close projected separations.

The results we obtain from the tailored-control method is largely consistent with the mass-binning method. Between the two methods we see that the $\Delta$log(sSFR) profile is independent of the total stellar mass of the galaxies. In the next two sections, we will use the tailored-control method to study how the difference profiles behave as a function of the mass ratio (\S~\ref{sec:dm}) and projected separation (\S~\ref{sec:sep}).

\subsection{Dependency on Mass Ratio}\label{sec:dm}

\begin{figure*}
\centering
\includegraphics[width=0.8\linewidth]{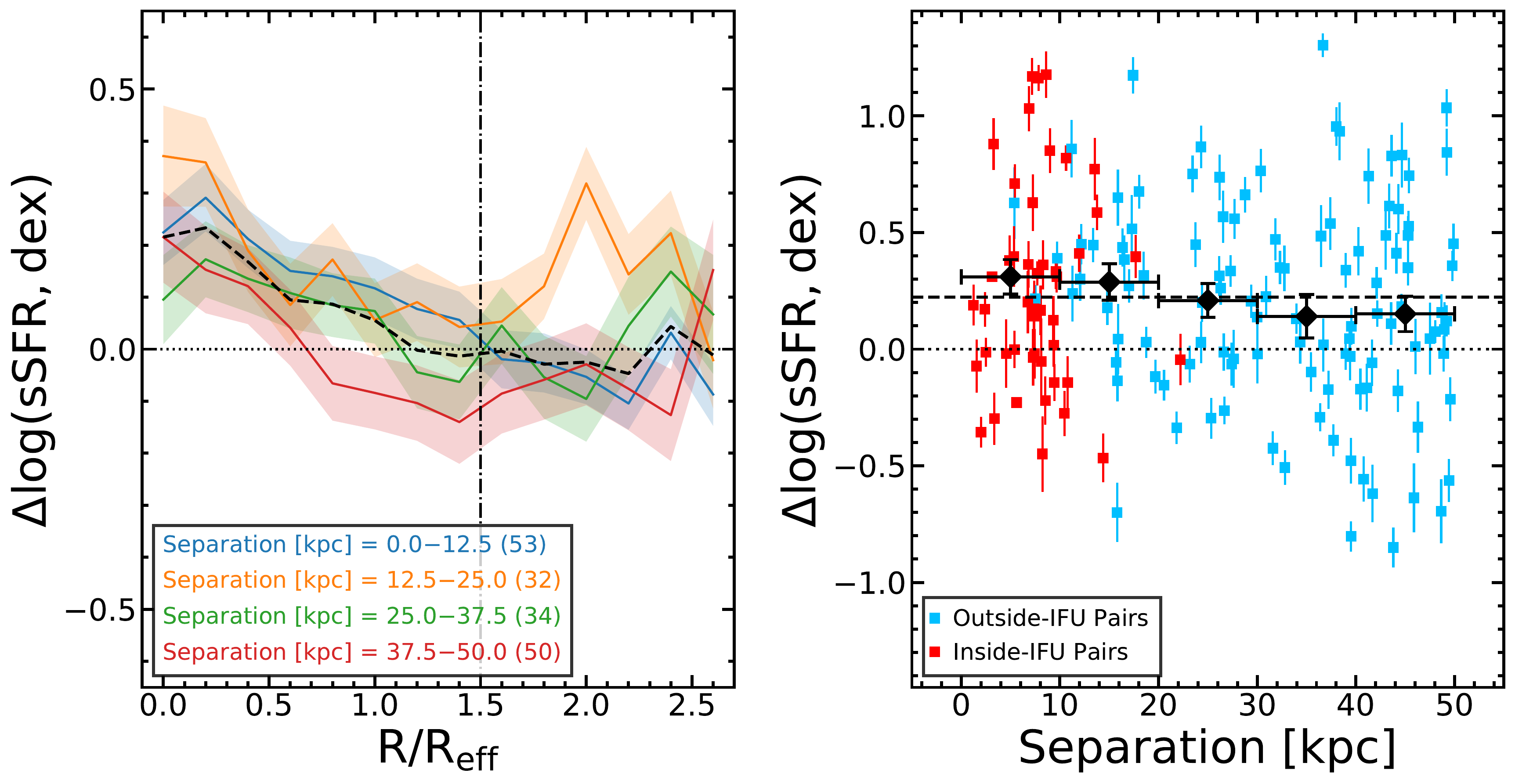}
\caption[]{Same as Figure \ref{fig:ssfr_mass} except the difference profiles are split by projected separation. }
\label{fig:ssfr_sep}
\end{figure*}

While the centrally enhanced star formation in close galaxy pairs seems mostly mass-independent, the mass ratio of the galaxy pair, like the projected separation, may be an important parameter that controls the level of enhancement \citep{Ellison:2008}. 

The mass ratio here is defined as, 
\begin{equation}\label{eq:dm}
\Delta {\rm log}(M) = {\rm log}(M_{\rm target}) - {\rm log}(M_{\rm comp}),
\end{equation}
where $M_{\rm target}$ is the stellar mass of the MaNGA target galaxy and $M_{\rm comp}$ is the stellar mass of its identified companion galaxy. In the inside-IFU sample, we have stellar masses for the MaNGA target galaxy but not the other identified pairs. Because of this, we leave the inside-IFU companions out of this analysis.

In the top left of Figure \ref{fig:ssfr_dm}, we split the $\Delta$log(sSFR) profiles into four mass ratio bins from $|\Delta$log(M)$|$ $=$ 0.0$-$1.0. Here, $|\Delta$log(M)$|$ $\le$ 0.5, represents equal mass major mergers while $|\Delta$log(M)$|$ $\ge$ 0.5 represents minor mergers. We see that there is a clear preference for pairs in major merger to have higher sSFR enhancements in their centers, $\sim$0.3$-$0.4 dex, while minor mergers show a weaker sSFR enhancement of $\sim$0.0$-$0.2 dex. The effect can be seen more clearly with the data extracted from the central 0.5 \reff\ in the top right panel. Here the $\Delta$log(sSFR) falls linearly with wider mass ratios, reaching zero enhancement to the sSFR at $|\Delta$log(M)$|$ $=$ 1.5$-$1.75.

We not only use the mass ratio to study how a major merger behaves differently from a minor merger but also to study how the more massive galaxy of a pair behaves differently from the less massive galaxy of the pair. We expect that either component of a similar mass pair will behave in a similar fashion; however, in a minor merger the large central galaxy may not respond to the merger event in the same way as its less massive companion. The two different scenarios can be differentiated with our definition of the mass ratio in equation \ref{eq:dm}; when the MaNGA target is the more massive companion of the pair, its mass ratio is positive and when MaNGA target is the less massive companion of pair the mass ratio is negative. To distinguish between these scenarios, we refer to the more massive galaxy of a pair as the primary companion and the less massive galaxy of a pair as the secondary companion.

In bottom left panel of Figure \ref{fig:ssfr_dm}, we split the profiles by mass ratio in four bins. The primary companions are the pairs with $\Delta$log(M) $\ge$ 0.0 and the secondary companions are the pairs with $\Delta$log(M) $\le$ 0.0. When splitting the mass ratio like this we see that major mergers still higher levels of sSFR enhancement in comparison to minor mergers; however, we also see that the less massive companion of the pair sees a higher $\Delta$log(sSFR) in comparison to the more massive companion of a pair. In major mergers, the sSFR enhancement is $\sim$0.1 dex higher in secondary companions in comparison to primary companions. In minor mergers the sSFR enhancement is $\sim$0.2 dex higher in secondary companions in comparison to primary companions.

In bottom right panel of Figure \ref{fig:ssfr_dm} we see that the sSFR enhancement is strongest in galaxies close to a 1:1 mass ratio ($\Delta$log($M$) $=$ 0.0). These galaxies feature a central enhancement of $\sim$0.4 dex which is 0.2 dex higher than the average central enhancement of the whole sample. This enhancement falls with wider mass ratios with a different slope for primary and secondary companions. At $\Delta$log($M$) $=$ -1.0, the sSFR is enhanced by 0.2 dex while at $\Delta$log($M$) $=$ 1.0 the sSFR enhancement is zero.

We also see that the $\Delta$log(sSFR) in the outskirts ($R$ $>$ 1.5 \reff) of the pairs have a dependence on the mass ratio in the bottom right of Figure \ref{fig:ssfr_dm}. Primary companions show a positive enhancement of 0.0$-$0.4 dex to the sSFR in their outskirts, while secondary companions feature a suppression to the sSFR in their outskirts of about 0.0$-$0.3 dex. Again we note that only 37\% of the MaNGA sample is designed to extend beyond 1.5 \reff; however, a similar result has been observed in the hydrodynamical simulations of \citet{Moreno:2015} and \citet{Moreno:2020}. 

From the mass ratio, we see two different effects. First, we see that the central enhancement to the sSFR is strongest for pairs with 1:1 mass ratios. Second, we see that secondary companions show steeper enhancement profiles with higher levels of sSFR enhancement in their centers and stronger levels of sSFR suppression in their disks with respect to primary companions. Finally, we see that primary companions feature sSFR enhancement at wide radii while secondary companions feature sSFR suppression at wide radii.

\subsection{Dependency on Projected Separation}\label{sec:sep}

A number of previous studies have shown that the sSFR enhancement increases as projected separation decreases \citep[e.g.,][]{Li:2008, Ellison:2008, Scudder:2012, Patton:2013}, in agreement with simulations \citep{Scudder:2012}. 

Figure \ref{fig:ssfr_sep} shows the profile and the central level of sSFR enhancement as a function of the projected separation ($r_{\rm p}$) from our MaNGA data. The $\Delta$log(sSFR) profiles show only a weak dependency on the projected separation. The pairs below a separation of 25 kpc have $\Delta$log(sSFR) profiles which lie $\sim$0.1 dex above the sample median while the profiles of the pairs beyond 25 kpc are $\sim$ 0.1 dex below the sample median. 

A dependence on the projected separation can be seen more clearly when looking at the data extracted from the inner 0.5 \reff. The level of the enhancement gradually increases with closer separation from 50 kpc to 10 kpc. While $\Delta$log(sSFR) falls at higher separations, there is still a substantial level of enhancement between 40 and 50 kpc, $\sim$0.15 dex. Within 10 kpc, the $\Delta$log(sSFR) enhancement jumps to $\sim$0.3 dex. Our data shows that the sSFR enhancement persists out to at least $r_{\rm p} = 50$~kpc, which is the limit of our pair selection.

\section{Discussion}\label{sec:disc}

\subsection{Radial Profiles of Enhancement}

\begin{figure}
\centering
\includegraphics[width=3in]{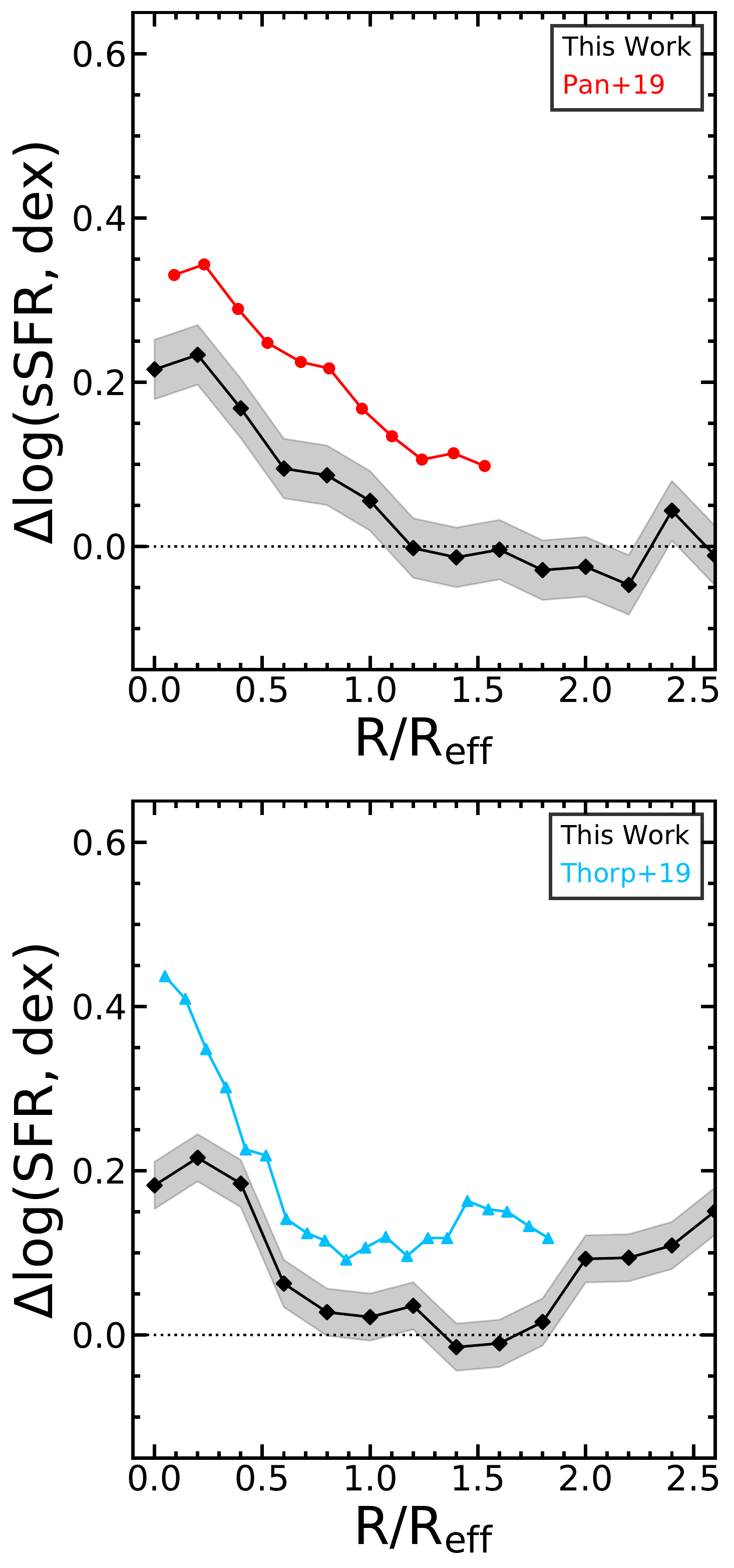}
\caption[]{The mean radial profile of $\Delta$log(sSFR) (Top) and $\Delta$log(SFR) (Bottom) between pairs and controls of this work with the tailored control method ({\it Black}) compared against those of \citet{Pan:2019} ({\it Red}), a pair sample, and \citet{Thorp:2019} ({\it Blue}), a post-merger sample.}
\label{fig:prof_comp}
\end{figure}

In Figure \ref{fig:prof_comp} we compare the $\Delta$log(sSFR) profile and $\Delta$log(SFR) as a function of galactocentric radius between our work and previous works using the MaNGA data. \citet{Thorp:2019} studies a sample of 36 post-mergers from the MaNGA sample. The centers of the post-merger galaxies feature a greater level of sSFR enhancement in their centers compared to our galaxies pairs of $\Delta$log(SFR) $=$ 0.40$-$0.45, roughly 0.2 dex higher than our paired galaxies. Between 0.4$-$1.0\reff\ the post-merger galaxies and our pairs are in closer agreement, being 0.05$-$0.1 dex higher than our paired galaxies. Beyond 1.0 \reff\ the post-merger galaxies have a higher $\Delta$log(SFR) by about 0.1$-$0.2 dex. The heightened level of sSFR enhancement in the centers of post-merger galaxies in comparison to merging galaxies seems consistent with the idea that a second and greater burst of star formation occurs as the two merging galaxies coalesce into a single galaxy. 

\citet{Pan:2019} studies a set of 205 paired galaxies in the MaNGA sample, focusing on the sSFR enhancement as a function of the merger interaction stage. The $\Delta$log(sSFR) profiles of the galaxies pairs in \citet{Pan:2019} are $\sim$0.10$-$0.15 dex above the profiles in our sample. Our paired galaxies show a slight sSFR suppression in their disks past 1.0 \reff\ while the galaxy pairs in \citet{Pan:2019} show an enhancement of $\sim$0.15 dex to the sSFR in their disks. 

The reason for the discrepancies between our samples may be that \citet{Pan:2019} includes a set of 96 pairs which are too close to be separately de-blended by SDSS and instead were visually identified by their morphology. 82 of these pairs are post-merger galaxies. \citet{Pan:2019} showed that pairs with disturbed morphology have higher $\Delta$log(SFR) when compared to pairs without morphological distortions. Further, \citet{Thorp:2019} showed that post-merger galaxies have $\Delta$log(SFR) profiles which are elevated over our sample's profiles. From this, infer that the lower $\Delta$log(sSFR) seen in our sample is due to our sample's lack of post-merger galaxies. 

\subsection{Central Enhancement vs. Merger Parameters}

While in this work we find that $\Delta$log(sSFR) has essentially no dependence on the stellar mass of the galaxy, \citet{Li:2008} found that lower mass galaxies experience greater levels of enhancement than higher mass galaxies. Galaxies in the mass range \logm\ $\ge$ 9.72 were shown to experience an enhancement $\sim$0.4 dex higher than galaxies in the mass range \logm\ $\ge$ 10.6. 

\begin{figure}
\centering
\includegraphics[width=3in]{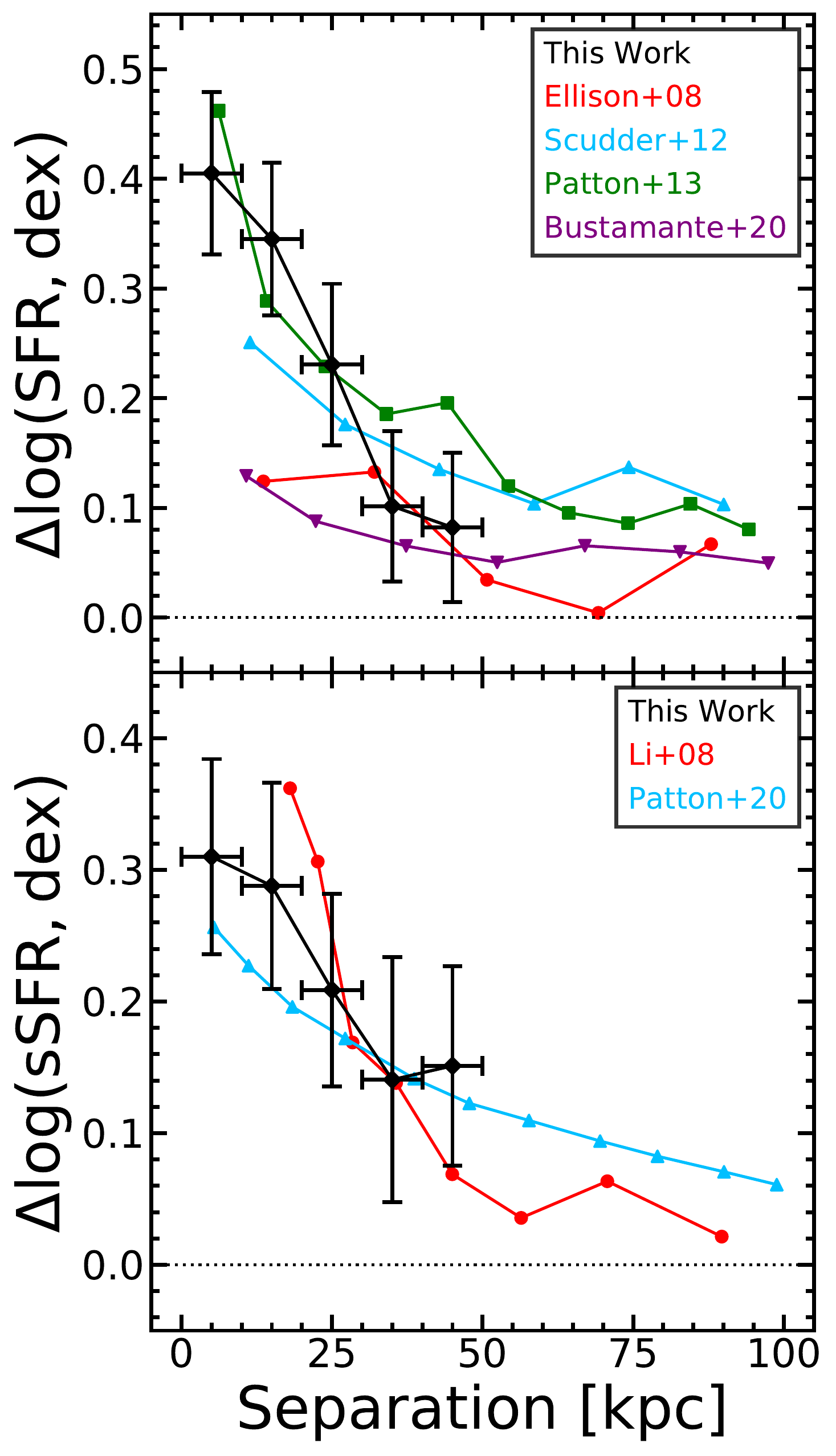}
\caption[]{(Top) $\Delta$log(SFR) extracted from the inner 0.5 \reff over projected separation from this study ({\it Black}), \citet{Ellison:2008} ({\it Red}), \citet{Scudder:2012} ({\it Blue}), \citet{Patton:2013} ({\it Green}), and \citet{Bustamante:2020} ({\it Purple}). (Bottom) $\Delta$log(sSFR) extracted from the inner 0.5 \reff over projected separation from this study ({\it Black}), \citet{Li:2008} ({\it Red}), and \citet{Patton:2020} ({\it Blue}).
}
\label{fig:nuc_sep}
\end{figure}

The source of the difference is unclear; between our works we utilize different methods for determining the sSFR enhancement in galaxy pairs. \citet{Li:2008} defines their enhancement function as the average sSFR of the paired galaxies at a given separation, weighted by the number of companions, subtracted by the average sSFR of the whole sample. While we do not find the same dependency between the sSFR enhancement and stellar mass, the sSFR enhancement as a function of projected separation between our studies is in good agreement (see Figure~\ref{fig:nuc_sep}).

The dependency of the SFR enhancement on the mass ratio was studied in \citet{Ellison:2008}. They found that pairs with mass ratios of 2:1 have higher SFR enhancements of $\sim$0.1 dex over pairs with wider 10:1 mass ratios. We find that pairs with mass ratios of $\Delta$log($M$) $=$ 0.25 (1.8:1) are 0.2 dex over pairs with mass ratios of $\Delta$log($M$) $=$ 1.0 (10:1). \citet{Ellison:2008} also saw tentative evidence for SFR enhancement in the secondary companions of minor pairs. We not only confirm that secondary companions of minor pairs feature SFR enhancement, but also that the enhancement is higher than the primary component of pairs at the same mass ratios. 

The observations are also consistent with simulations. \citet{Moreno:2015} used {\sc Gadget}-3, a smoothed particle hydrodynamics code, to study the spatial extent of the star formation enhancement. The merger simulations showed that the star formation enhancement was largely concentrated in the centers (R $<$ 1 kpc) of the paired galaxies and that the off-nuclear regions (1 $<$ R $<$ 10 kpc) showed a suppression to the star formation. \citet{Moreno:2015} also found that lower mass secondary galaxies (in mergers with stellar mass ratios of 2.5:1) have higher levels of sSFR enhancement in their centers and have stronger levels of sSFR suppression in their disks. This is in good agreement with our work, we found that secondary companions show moderately higher levels of sSFR enhancement in their centers in comparison to primary companions at the same mass ratio. Further, the outskirts of our primary companions feature positive enhancement to their sSFR while the secondary companions show a strong suppression to the sSFR in their outskirts. 

Figure~\ref{fig:nuc_sep} compares the enhancement in both SFR and sSFR as a function of projected separation. For our data points, we use the mean $\Delta\log{\rm SFR}$ and $\Delta\log{\rm sSFR}$ measured within a deprojected radius of 0.5~\reff\ (i.e., the black data points in the right side panel of Figure~\ref{fig:ssfr_sep}). The literature data is measured from single 1-1.5\arcsec-radius SDSS fibers \citep{Ellison:2008,Scudder:2012,Patton:2013,Bustamante:2020}. We find that our central SFR enhancements for closely separated pairs are higher than many of the previous studies, our sSFR enhancement is $\sim$0.1 dex higher than \citet{Scudder:2012} and $\sim$0.2 dex higher than \citet{Ellison:2008} and \citet{Bustamante:2020}. The central enhancement to the sSFR for closely separated pairs in our sample is $\sim$0.05 dex lower than \citet{Patton:2013}. The sSFR enhancement function from \citet{Li:2008} is 0.05$-$0.10 dex above our sample at close separations.   

The sSFR as a function of separation was studied with the cosmological hydrodynamical simulations of IllustrisTNG \citep{Patton:2020}. In Figure \ref{fig:nuc_sep} we show the $\Delta$log(sSFR) enhancements across 2D separation (projected separation) from the TNG300-1 simulation. \citet{Patton:2020} found a sSFR enhancement which is 1.8$\times$ that of the isolated controls within a separation of 15 kpc and is statistically significant out to a 2D separation of $\sim$250 kpc. This enhancement is $\sim$0.05 dex lower than ours; however, our sSFR enhancement shown in Figure \ref{fig:nuc_sep} was extracted from a 0.5 \reff\ aperture while \citet{Patton:2020} extracted the sSFR from a 50\% half-mass radius. This means that our extraction radius is effectively twice a small and is more restricted to the central burst of star formation. 

Our sample only covers projected separations within 50 kpc, while previous surveys cover out to 100$-$200 kpc. While our sample covers a smaller separation range, the $\Delta$log(SFR) of our sample at 50 kpc is roughly the same level as \citet{Ellison:2008} and \citet{Bustamante:2020} at the same projected separation. From this we see that our sSFR enhancement as a function of projected separation is consistent with with what has been found in previous works. 

\subsection{Comparison with Simulations}

Our results are consistent with the idea that galaxy merger events trigger a burst of star formation in the centers of the paired galaxies. In this model, when two galaxy pairs undergo their first pericenter, the tidal torques exerted on the disks of interacting galaxies form barred structures. These barred structures funnel gases into the centers of the galaxies which triggers a burst of new star formation. Eventually, as the pairs separate from each other after the first pericenter, the burst of star formation begins to subside. This is shown in our work and many previous works in Figure \ref{fig:nuc_sep} where the sSFR enhancement is greatest for close separations and falls with wider separations.

We also found that the $\Delta$log(sSFR) is independent of the stellar mass. This means that the merger event will produce the same amount of new stellar mass for low mass galaxies as it produces for high mass galaxies. This also means that the low mass galaxies will experience a greater change in total stellar mass before and after the merger event so the merger event will have a greater impact on the mass evolution of low mass galaxies as opposed to high mass galaxies.

We find that the strength of the central burst of merger induced star formation is dependent on the relative masses between the two galaxies. Equal mass galaxies see the strongest bursts of star formation while wide mass ratios see weaker bursts of star formation. We further find that the strength of the central burst of star formation differs between the primary and secondary companion of a pair where the less massive secondary companion features a higher level of sSFR enhancement in comparison to its higher mass primary companion. 

\citet{Moreno:2020} uses galaxy merger simulations to study the origin of the enhanced star formation in galaxy pairs, whether its an increase in the star formation efficiency, defined as the ratio between the SFR and the mass of cold-dense gas, or its an increase in the availability of cold dense gas. They find that the star formation in secondary galaxies is evenly split between being efficiency driven or fuel driven systems while primary galaxies are more likely to be fuel driven systems (71\%). \citet{Moreno:2020} also finds that that secondary galaxies feature higher levels of star formation enhancement in comparison to primary galaxies, which is in agreement with what we find in this work. This indicates that the reason why primary galaxies behave differently from secondary galaxies may be due to a difference in the physical mechanism which drives the enhanced star formation. 

We also find that there is a difference in the $\Delta$log(sSFR) offsets at wide radii between primary and secondary companions. The primary companion features an enhancement to their sSFR at wide radii while the secondary companion features a suppression to their sSFR at wide radii. The reason for this is unclear; however, this means that the primary companions will experience enhanced stellar mass growth across their disks while the secondary companions will see only see substantial stellar mass growth in their centers which will result in more bulge dominated galaxies. Differences in the bulge-to-total ratio, (B/T), has been observed in previous works like \citet{Bluck:2019} who found that satellites tend to have slighter higher (B/T) ratios in comparison to central galaxies at the same stellar mass.

\section{Summary and Conclusion}\label{sec:sum}

In this paper, we have demonstrated the power of a massive integral-field spectroscopic survey in comparison studies of galaxy properties. The nearby galaxy populations show consistent behaviors despite of their large diversity in star formation properties. Stacking was able to detect the signal buried in the noise by averaging, in the time domain, the likely stochastic star forming histories of galaxies (which drives the scatter in the sample). We focused on comparing the azimuthally averaged radial profiles of sSFR between galaxies in close pairs and a control sample of isolated galaxies. In agreement with previous studies, we found that, on average, star formation is elevated in close galaxy pairs. The properties of these purported merger-induced differences in sSFR can be summarized as follows:

\begin{enumerate}

\item Star formation is enhanced within the inner 1.0 \reff\ and it peaks at a level of 0.20$-$0.25 $\pm$ 0.1 dex (i.e., $\sim$2$\times$ faster star formation). On the other hand, the outskirts of the paired galaxies (\reff\ $=$ 1.0$-$2.5) show a moderate amount of suppression in sSFR at a level of $\sim$0.1 dex.
 
\item The sSFR difference profile is largely independent of the stellar mass.

\item The level of central sSFR enhancement increases smaller projected separations. 

\item The level of central sSFR enhancement depends on the mass ratio of the galaxy pair. Galaxies with small mass ratios, $|\Delta$log($M$)$|$ $=$ 0.0$-$0.5, see an enhancement $\sim$0.1$-$0.2 dex higher than the average. But the enhancement is still present in pairs with mass ratios as large as $|\Delta$log($M$)$|$ $=$ 1.0. 

\item The merger-induced changes in sSFR also seem to differ between the more massive and the less massive member of a galaxy pair. At the same mass ratio, the less massive member in a galaxy pair shows a higher sSFR enhancement (by $\sim$0.1$-$0.2 dex) in comparison to the more massive member. 

\item At large radii (R/\reff\ $>$ 1.5), the more massive companion of pairs show an enhancement to the sSFR while the less massive companion of pairs show a suppression to the sSFR.

\end{enumerate}

As of 2020 August 24, MaNGA has completed its observations of $\sim$10,000 galaxies. This final sample gives us access to an unprecedentedly large sample of paired and control galaxies with 2\arcsec\ spatial sampling up to 2-3~\reff, which will undoubtedly improve the results presented here. In addition, the less massive members of the inside-IFU pairs can be included in such a comparison study once we build a better deblended photometric catalog using SDSS images.

\acknowledgments

We thank the anonymous referee for the useful comments and suggestions. We also thank Cheng Li for our useful discussion on how the sSFR enhancement in galaxy pairs relates to their stellar mass and mass ratio. J.S. and H.F. acknowledge support from the National Science Foundation (NSF) grant AST-1614326. YSD would like to acknowledge support from NSFC grant number 10878003. Funding for the Sloan Digital Sky Survey IV has been provided by the Alfred P. Sloan Foundation, the U.S. Department of Energy Office of Science, and the Participating Institutions. SDSS acknowledges support and resources from the Center for High-Performance Computing at the University of Utah. The SDSS web site is www.sdss.org.

SDSS is managed by the Astrophysical Research Consortium for the Participating Institutions of the SDSS Collaboration including the Brazilian Participation Group, the Carnegie Institution for Science, Carnegie Mellon University, the Chilean Participation Group, the French Participation Group, Harvard-Smithsonian Center for Astrophysics, Instituto de Astrofísica de Canarias, The Johns Hopkins University, Kavli Institute for the Physics and Mathematics of the Universe (IPMU) / University of Tokyo, the Korean Participation Group, Lawrence Berkeley National Laboratory, Leibniz Institut für Astrophysik Potsdam (AIP), Max-Planck-Institut für Astronomie (MPIA Heidelberg), Max-Planck-Institut für Astrophysik (MPA Garching), Max-Planck-Institut für Extraterrestrische Physik (MPE), National Astronomical Observatories of China, New Mexico State University, New York University, University of Notre Dame, Observatório Nacional / MCTI, The Ohio State University, Pennsylvania State University, Shanghai Astronomical Observatory, United Kingdom Participation Group, Universidad Nacional Autónoma de México, University of Arizona, University of Colorado Boulder, University of Oxford, University of Portsmouth, University of Utah, University of Virginia, University of Washington, University of Wisconsin, Vanderbilt University, and Yale University.

\end{CJK*} 
\bibliographystyle{aasjournal}
\bibliography{mergerbib}

\end{document}